\input{aipcheck}

\documentclass[
    ,final            
  ]
  {aipproc}

\layoutstyle{8x11double}

\begin{document}

\title{Phantom collapse of electrically charged scalar field\\ in dilaton gravity}


%
%
\author{Anna Nakonieczna\footnote{aborkow@kft.umcs.lublin.pl} and~Marek Rogatko\footnote{rogat@kft.umcs.lublin.pl, marek.rogatko@poczta.umcs.lublin.pl}}
{address={Institute of Physics \\
Faculty of Mathematics, Physics and~Computer Science \\
Maria~Curie-Sk{\l}odowska~University \\
20-031 Lublin, pl. Marii Curie-Sklodowskiej 1, Poland}
}

\begin{abstract}
Our research focus on gravitational collapse of electrically charged scalar field in dilaton gravity and~in the presence of phantom coupling. We examine dynamical behaviour of the scalar field coupled to Maxwell field when gravitational interactions have form consistent with the low-energy limit of the string theory. Moreover, we allow the evolving fields to~have negative sign in front of the respective kinetic term of the Lagrangian. The main aim of our studies is to investigate in~what manner does the phantom nature of either Maxwell or dilaton fields (or both of them) affect the outcomes of~the~collapse. It~turns out that the influence is crucial to the obtained spacetime structures. Negative kinetic energy of one (or both) of~the~fields delays, changes the course or even prevents the collapse.
\end{abstract}

\maketitle


\section{Introduction}

Dark energy is a~highly predominating constituent of the Universe. Hence, an interesting issue arises: in what manner does its presence affect the processes, which are driven by gravitational force?
In order to answer this question the properties of dark energy ought to be unraveled. Although it seems to be so widespread, the problem of its exact nature is still unresolved. There are numerous diverse theoretical models which attempt to describe and~explain the properties of the considered unknown constituent of the Universe.

Dark energy is described by the equation of state \mbox{$w=P\rho^{-1}$}, where $P$ and~$\rho$ correspond to its pressure and~density, respectively~\cite{AmendolaTsujikawa}. Although the exact value of the barotropic index $w$ is still unknown, observations do not exclude the possibility of $w<-1$~\cite{CaldwellKamionkowski2009-397}. In this particular case, the so-called phantom fields provide a~theoretical model for dark energy~\cite{Caldwell2002-23,CaldwellKamionkowskiWeinberg2003-071301,DabrowskiStachowiakSzydlowski2003-103519}. Our research concern the influence of phantom nature of evolving fields on dynamical gravitational collapse. We are particularly interested in simulating the collapse of electrically charged scalar field, which allows us to draw conclusions about outcomes of the realistic process taking place in nature.

Astrophysical black holes are electrically neutral and~rotating~\cite{Ori1992-2117}. The non-zero angular momentum results in axial symmetry of these objects. Within the framework of the general theory of relativity the structure of spacetime containing such a~realistic black hole is described by the Kerr metric~\cite{FrolovNovikov,Chandrasekhar}. Unfortunately, investigating gravitational collapse which leads to the formation of the~Kerr black hole is extremely difficult due to analytical and~numerical obstacles. For this reason some simplifications ought to be introduced. Basing on striking similarities in causal structures of spacetimes containing Kerr and~Reissner-Nordstr\"{o}m black holes the latter spacetime may be regarded as a~toy model for the former one~\cite{FrolovNovikov,Yurtsever1993-17}. Since it is spherically symmetric the calculations simplify considerably. The dynamical Reissner-Nordstr\"{o}m spacetime stems from the collapse of electrically charged scalar field~\cite{OrenPiran2003-044013}. Hence spacetime structure emerging from the evolution of such a~field actually imitates a~structure resulting from the realistic gravitational collapse.

We simulate the dynamical behaviour of complex scalar field coupled to Maxwell field when gravitational interactions take form of dilaton gravity and~phantom coupling of Maxwell and~dilaton fields is possible. We~are primarily interested in singular spacetimes since examining the dynamical evolution of fields under the influence of gravity allows us to describe internal structures of the objects contained within them properly.


\section{Einstein-Maxwell-dilaton theory with phantom coupling}

The action, which allows us to investigate gravitational collapse in the considered theory, represents complex scalar field coupled to Maxwell field in dilaton gravity. Moreover, it takes phantom coupling of Maxwell and~dilaton fields into account. In the {\it string frame} it has the following form:
\begin{eqnarray}
S^{_{\left(SF\right)}} = \int d^4x \sqrt{-g^{_{\left(SF\right)}}}
e^{-2\phi}\Big[R^{_{\left(SF\right)}}-2\xi_1\big(\nabla^{_{\left(SF\right)}}\phi\big)^2 + \nonumber \\
+e^{2\alpha\phi}\mathcal{L}^{_{\left(SF\right)}}\Big],
\end{eqnarray}
where $\mathcal{L}^{_{\left(SF\right)}}$ stands for Lagrangian density
\begin{eqnarray}
\mathcal{L}^{_{\left(SF\right)}} = -\frac{1}{2} D_\mu\psi D^\mu\psi^\ast -\xi_2F_{\mu\nu}F^{\mu\nu},
\end{eqnarray}
while $\phi$ is the dilaton field, $\psi$ -- complex scalar field, $F_{\mu\nu}\equiv A_{\nu,\mu}-A_{\mu,\nu}$ stands for the electromagnetic field tensor, where $A_\mu$ is the Maxwell field. The covariant derivative of the scalar field is defined as \mbox{$D_\mu\equiv\nabla^{_{\left(SF\right)}}_\mu+ieA_\mu$}. Coupling constants $e$ and~$\alpha$ characterize couplings between the scalar field and~Maxwell and~dilaton fields, respectively. Phantom constants $\xi_1$ and~$\xi_2$ refer to dilaton and~Maxwell fields, respectively, and~are equal to $+1$ or $-1$. In the latter case the respective field is phantom.

The equations of motion for the considered fields are obtained via~variational principle. During their derivation the {\it string frame} is exchanged for the {\it Einstein frame} according to conformal transformation \mbox{$g^{_{\left(EF\right)}}_{\mu\nu}=e^{-2\phi}g^{_{\left(SF\right)}}_{\mu\nu}$}. Due to the fact that the evolving fields are massless, the collapse is simulated in double null coordinates and~the assumed line element has the form \mbox{$ds^2=-a\left(u,v\right)^2dudv+r\left(u,v\right)^2d\Omega^2$}, where $a\left(u,v\right)$ is an arbitrary function, $r\left(u,v\right)$ is the radial function and~$d\Omega^2$ stands for the line element on the unit sphere.

Einstein equations obtained in accordance with the above assumptions and~conditions are given by
\begin{eqnarray}
\label{first}
\frac{2a_{,u} r_{,u}}{a} - r_{,uu} = \xi_1r\phi_{,u}^2 + \frac{1}{4} r e^{2\phi \left( \alpha~+ 1 \right)} \Big[ \psi_{,u} \psi_{,u}^\ast + \nonumber \\
+ ieA_u \left( \psi \psi_{,u}^\ast - \psi_{,u} \psi^\ast \right) + e^2 A_u^2 \psi \psi^\ast \Big], \\
\frac{2a_{,v} r_{,v}}{a} - r_{,vv} = \xi_1r\phi_{,v}^2 + \frac{1}{4} r e^{2\phi \left( \alpha~+ 1 \right)} \psi_{,v} \psi_{,v}^\ast, \\
r_{,uv} + \frac{a^2}{4r} + \frac{r_{,u} r_{,v}}{r} = \xi_2e^{2\alpha~\phi} \frac{Q^2 a^2}{4r^3}, \\
\frac{a_{,u} a_{,v}}{a^2} - \frac{a_{,uv}}{a} - \frac{r_{,uv}}{r} = \xi_1\phi_{,u} \phi_{,v} + \xi_2e^{2\alpha~\phi} \frac{Q^2 a^2}{4r^4} + \nonumber \\
+ \frac{1}{8} e^{2\phi \left( \alpha~+ 1 \right)} \Big[ \psi_{,u} \psi_{,v}^\ast + \psi_{,v} \psi_{,u}^\ast + \nonumber \\
+ ieA_u \left( \psi \psi_{,v}^\ast - \psi_{,v} \psi^\ast \right) \Big].
\end{eqnarray}
The equations of motion of the complex scalar field are as follows:
\begin{eqnarray}
r\psi_{,uv} + r_{,u} \psi_{,v} + r_{,v} \psi_{,u} + ierA_u \psi_{,v} + ier_v A_u \psi + \nonumber \\
+ ie\psi \frac{Q a^2}{4r} = 0, \\
r\psi_{,uv}^\ast + r_{,u} \psi_{,v}^\ast + r_{,v} \psi_{,u}^\ast - ierA_u \psi_{,v}^\ast - ier_v A_u \psi + \nonumber \\
- ie\psi^\ast \frac{Q a^2}{4r} = 0,
\end{eqnarray}
while the equation of motion of the dilaton field has the following form:
\begin{eqnarray}
r\phi_{,uv} + r_{,v} \phi_{,u} + r_{,u} \phi_{,v} - \alpha~\frac{\xi_2}{\xi_1} e^{2\alpha~\phi} \frac{Q^2 a^2}{4r^3} + \nonumber \\
- \frac{\alpha~+ 1}{8\xi_1} r e^{2\phi \left( \alpha~+ 1 \right)} \Big[ \psi_{,u} \psi_{,v}^\ast + \psi_{,v} \psi_{,u}^\ast + \nonumber \\
+ ieA_u \left( \psi \psi_{,v}^\ast - \psi_{,v} \psi^\ast \right)\Big] = 0.
\end{eqnarray}
On account of gauge freedom of electromagnetic potential $A_\mu\rightarrow A_\mu+\nabla_\mu\Lambda$, where $\Lambda$ is an arbitrary scalar function, Maxwell equations may be written as
\begin{eqnarray}
Q \equiv \frac{2r^2}{a^2} A_{u,v}, \\
Q_{,v} + 2\alpha~\phi_{,v} Q + \frac{1}{4\xi_2} e^{2 \phi} ier^2 \left( \psi_{,v} \psi^\ast - \psi \psi_{,v}^\ast \right) = 0.
\label{last}
\end{eqnarray}

The set of equations (\ref{first})--(\ref{last}) describes the gravitational collapse in Einstein-Maxwell-dilaton theory with phantom coupling. Because of its complicated structure it has to be solved numerically. The suitable algorithm was described in~\cite{BorkowskaRogatkoModerski2011-084007}.

\section{Spacetime structures}

The initial conditions for analysed evolutions consist of profiles for complex scalar field and~dilaton field. They were chosen as trigonometric type with an amplitude $\tilde{p}_\psi$ and~gaussian type with an amplitude $\tilde{p}_\phi$, respectively~\cite{NakoniecznaRogatkoModerski2012-044043}. Since the dynamical gravitational collapse is universal, its outcomes do not depend on the specific form of initial profiles. They are also independent of the values of an electric coupling constant $e$ and~a~parameter $\delta$ provided that these are not equal to zero. For this reason, these two quantities were 
fixed in all the conducted evolutions.

On the other hand, it turns out that structures of spacetimes emerging from the collapse depend on initial amplitudes of the collapsing fields $\tilde{p}_\psi$ and~$\tilde{p}_\phi$ as well as on the value of dilatonic coupling constant $\alpha$. We considered two values of it, namely $\alpha=-1$ and~$\alpha=0$. The former refers to the low-energy string theory, while the latter causes the evolution to run in the presence of uncoupled dilaton field.

In order to examine the roles of evolving fields and~couplings among them in the considered gravitational collapse it is sufficient to describe its outcomes when initial amplitude of complex scalar field is constant and~the amplitude of dilaton field varies. An amplitude $\tilde{p}_\psi$ was set as equal to $0.6$ in all the evolutions.

The types of all phantom evolutions, which will be analysed, are characterized in Table \ref{tab:char}.

\subsection{Phantom Maxwell field}

\begin{figure}[h]
  \includegraphics[height=.475\columnwidth]{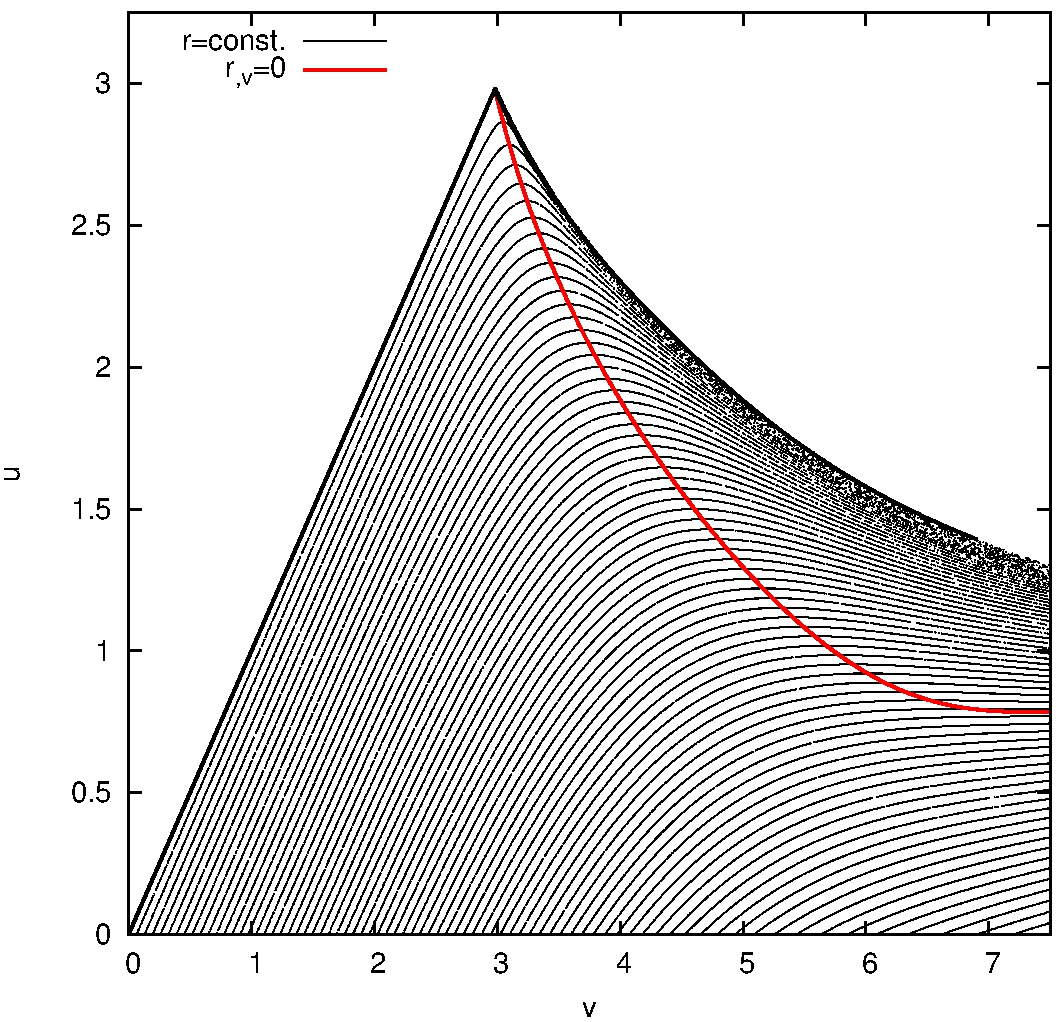}
  \caption{Dynamical singular spacetime emerging from the $E\overline{M}$-collapse.}
  \label{fig:E-M}
\end{figure}

Figure \ref{fig:E-M} presents the structure of singular spacetime emerging from the gravitational collapse of complex scalar field coupled to Maxwell field with phantom coupling to gravity, i.e. the result of an $E\overline{M}$-evolution. The singular spacetimes emerge in this case for large enough values of initial complex scalar field amplitudes. It turns out that the obtained structure corresponds to the dynamical Schwarzschild-type spacetime. An apparent horizon surrounding central spacelike singularity settles along $u=const.$ at $v\to\infty$ indicating the location of an event horizon. On the contrary, the outcome of the collapse of complex scalar field coupled to the non-phantom Maxwell field is the dynamical Reissner-\linebreak-Nordstr\"{o}m spacetime~\cite{OrenPiran2003-044013}. Apart from the central spacelike singularity surrounded by an apparent horizon coinciding with an event horizon at $v\to\infty$ it possesses a~Cauchy horizon at $v=\infty$.

Taking the above into account, it may be stated that phantom nature of Maxwell field prevents the formation of a~Cauchy horizon and~supports the emergence of the simplest spacetime structure containing a~black hole.

\begin{figure}[h]
  \includegraphics[height=.475\columnwidth]{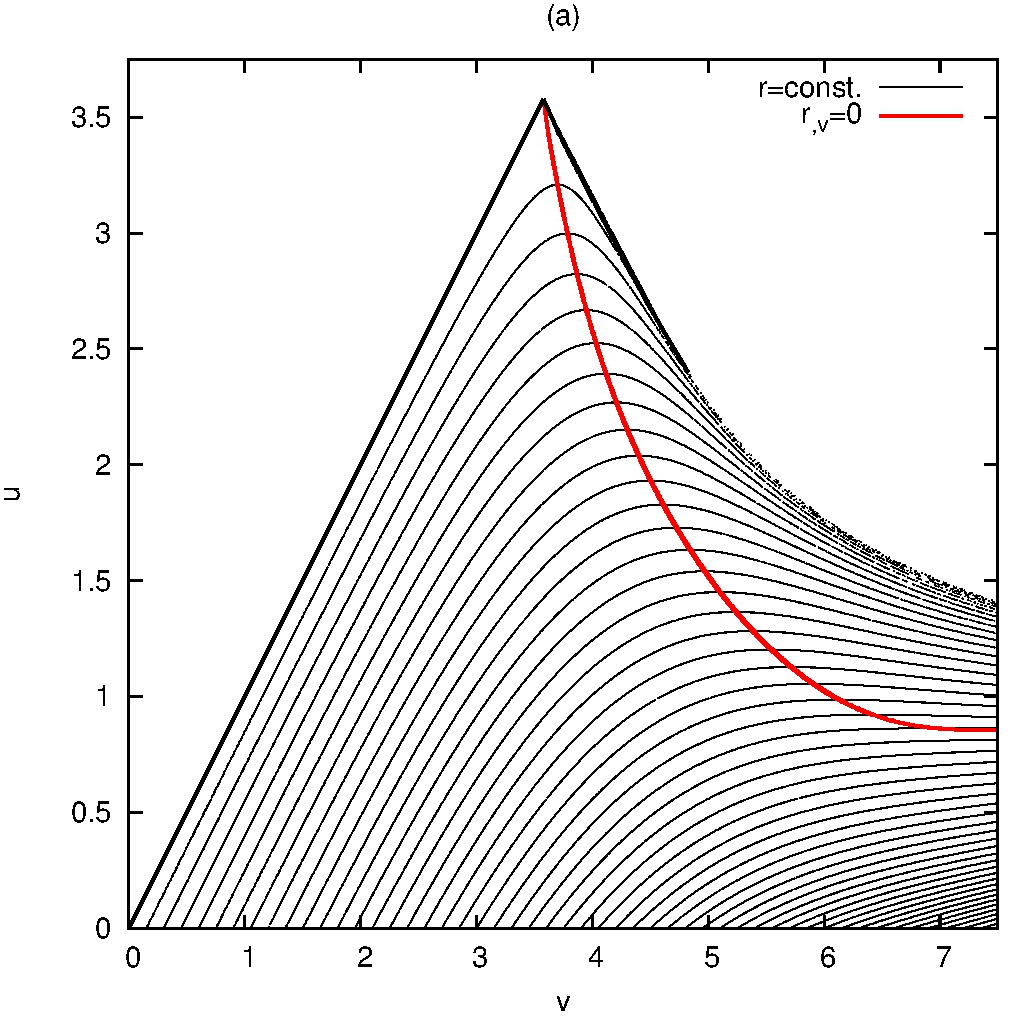}
  \includegraphics[height=.475\columnwidth]{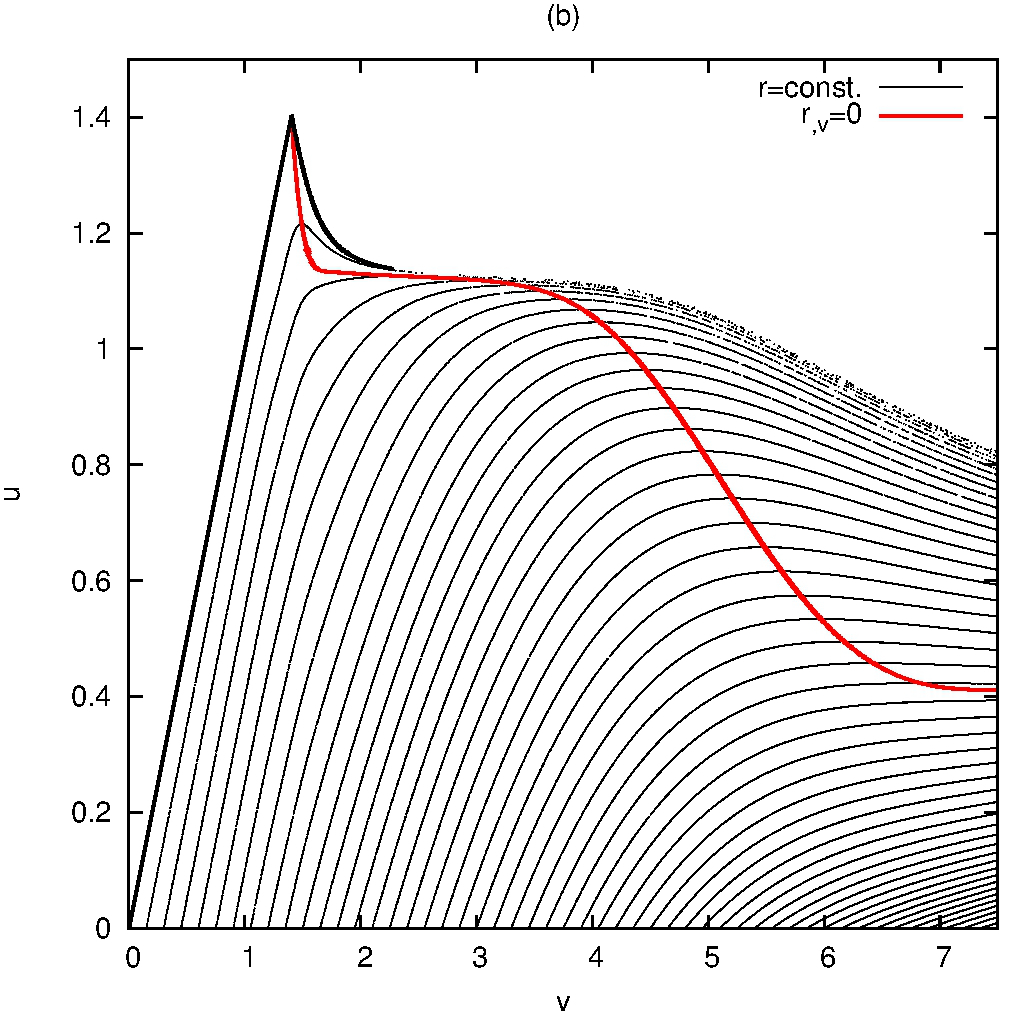}
  \caption{Dynamical singular spacetimes emerging from the $E\overline{M}D$-collapse for $\alpha=-1$ and~$\alpha=0$ when $\tilde{p}_\phi$ is equal to (a) $0.01$ and~(b) $0.09$.}
  \label{fig:E-MD}
\end{figure}

The results of an $E\overline{M}D$-collapse are identical for both considered values of $\alpha$. They are shown in Fig.\ref{fig:E-MD}. Regardless of the initial value of dilaton field amplitude all emerging spacetimes are singular and~Schwarzschild-\linebreak-type. For small values of the amplitude the spacetime structure is typical. For its bigger values the collapse runs in two stages, what results in an appearance of a~temporary horizon in spacetime. The joint collapse of phantom Maxwell field coupled to the complex scalar one and~non-phantom dilaton field leads to the formation of a~simplest spacetime structure describing a~black hole.

Since phantom Maxwell field favours the emergence of such a~structure, this result indicates that non-phantom dilaton field either also supports the process or is unable to counteract effectively.

\begin{table}[t]
\begin{tabular}{ccccc}
\hline
\tablehead{1}{c}{b}{Type}
  & \tablehead{1}{c}{b}{$\xi_1$}
  & \tablehead{1}{c}{b}{$\xi_2$}
  & \tablehead{1}{c}{b}{$\tilde{p}_\phi$}
  & \tablehead{1}{c}{b}{$\tilde{p}_\psi$}   \\
\hline
$E\overline{M}$ & $-$ & $-1$ & $-$ & $\neq 0$\\
$E\overline{M}D$ & $+1$ & $-1$ & $\neq 0$ & $\neq 0$\\
$E\overline{D}$ & $-1$ & $-$ & $\neq 0$  & $-$\\
$EM\overline{D}$ & $-1$ & $+1$ & $\neq 0$ & $\neq 0$\\
$E\overline{MD}$ & $-1$ & $-1$ & $\neq 0$ & $\neq 0$\\
\hline
\end{tabular}
\caption{Characteristics of the considered evolutions. Overlining indicates phantom coupling of the particular field.}
\label{tab:char}
\end{table}

\subsection{Phantom dilaton field}

The $E\overline{D}$-collapse represents the evolution of phantom scalar field under the influence of Einstein gravity. It~turns out that regardless of an initial value of the field's amplitude the resulting spacetime is non-singular. On account of the fact that dynamical Schwarzschild spacetime emerges from the collapse of non-phantom scalar field~\cite{HamadeStewart1996-497}, the obtained result means that phantom nature of the scalar field prevents singularity formation.

\begin{figure}[h]
  \includegraphics[height=.475\columnwidth]{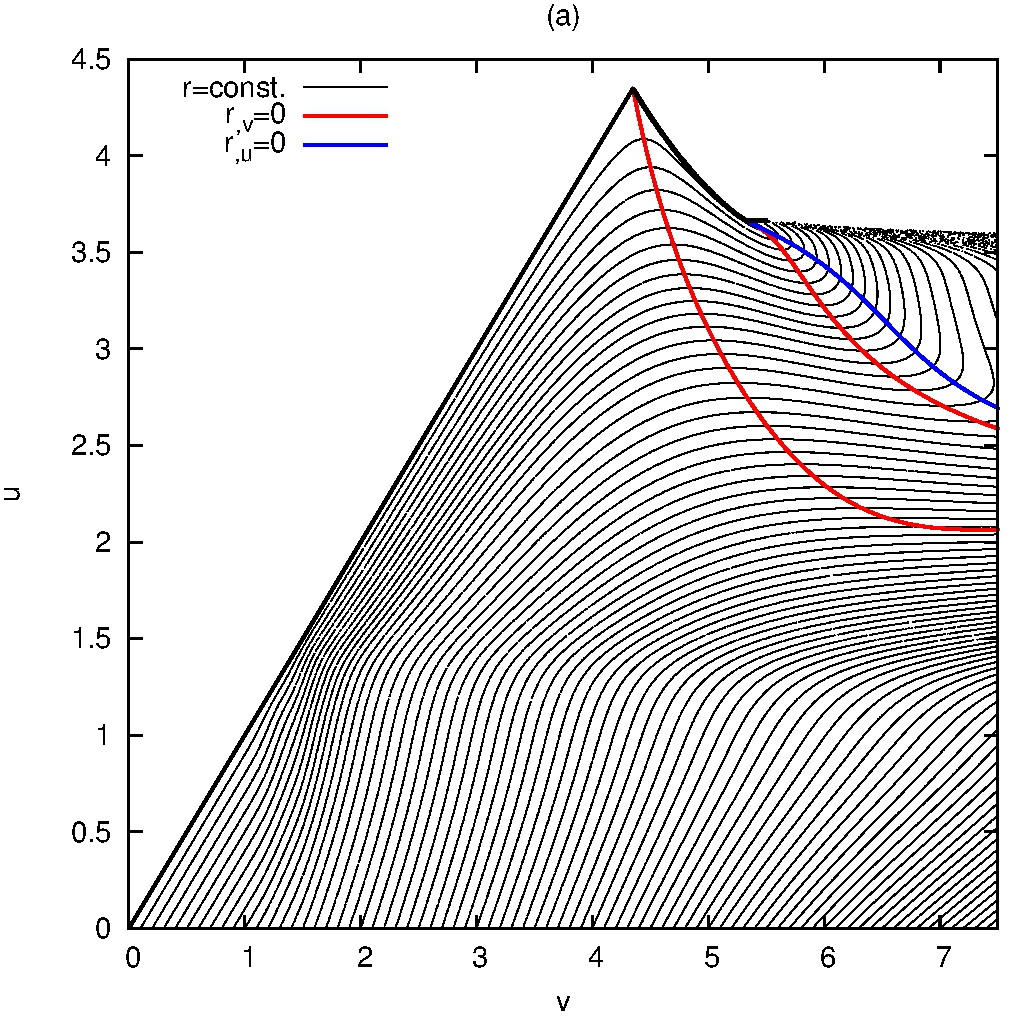}
  \includegraphics[height=.475\columnwidth]{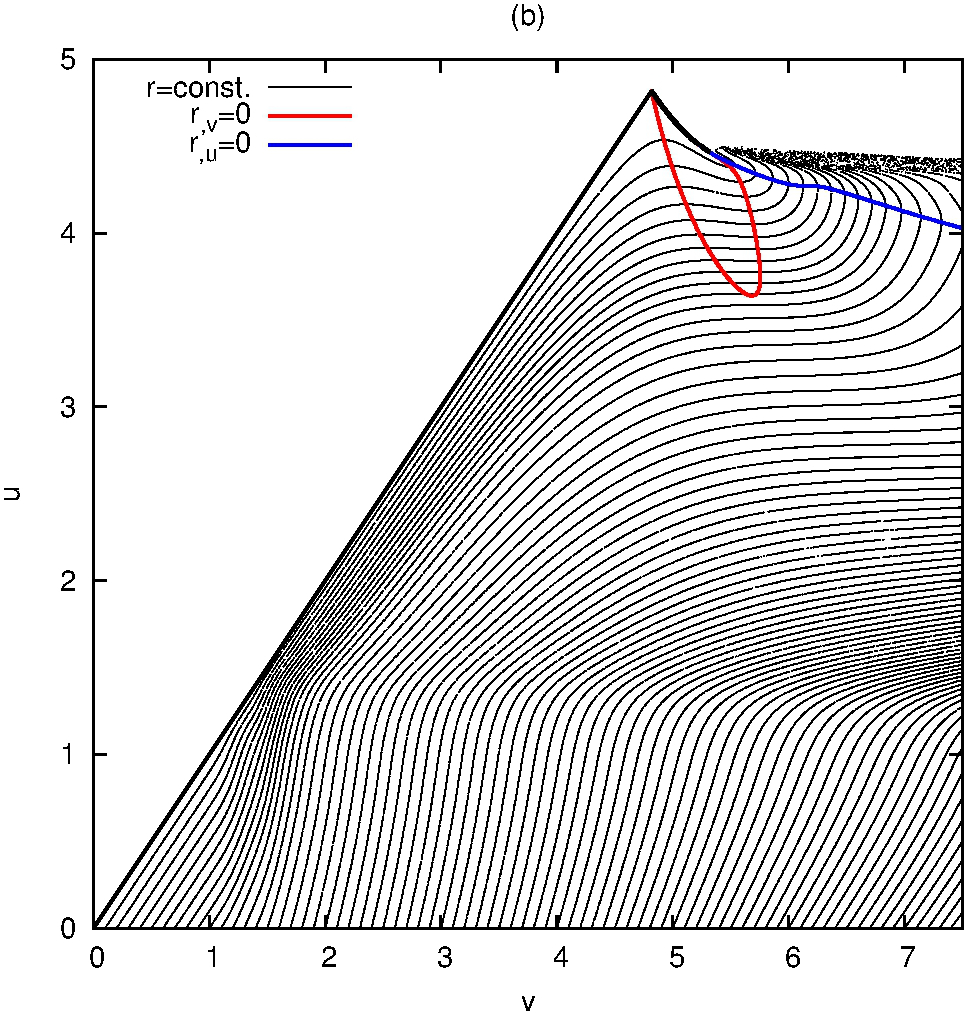}
  \caption{Dynamical singular spacetimes emerging from the $EM\overline{D}$-collapse for $\alpha=-1$ when $\tilde{p}_\phi$ is equal to (a) $0.15$ and~(b) $0.2$.}
  \label{fig:EM-D}
\end{figure}

A~collection of singular spacetime structures, which are obtained during the dynamical gravitational collapse in $EM\overline{D}$-theory for $\alpha=-1$, is presented in Fig.\ref{fig:EM-D}. For small values of dilaton field amplitude the initially formed central singularity, which is surrounded by an apparent horizon coinciding at $v\to\infty$ with the event horizon situated along $u=const.$, bifurcates and~forms two wormhole throats during the course of the collapse. Finally, there exists a~dynamical wormhole within the emerging spacetime~\cite{NakoniecznaRogatkoModerski2012-044043}. For bigger initial values of the amplitude of dilaton field the spacetime is singular, but the singularity is not surrounded by the event horizon, i.e. a~naked singularity forms. For even larger values of the dilaton field amplitude the resulting spacetime is\linebreak non-singular.

In contrast, the collapse of an electrically charged scalar field in the presence of phantom dilaton field when $\alpha=0$ does not lead to singular spacetimes.

Summarizing the above findings it may be stated that the tendency of phantom dilaton field to prevent the singularity formation is stronger in the uncoupled case, i.e. for $\alpha=0$, than for $\alpha=-1$.

\subsection{Phantom Maxwell and~dilaton fields}

\begin{figure}[h]
  \includegraphics[height=.475\columnwidth]{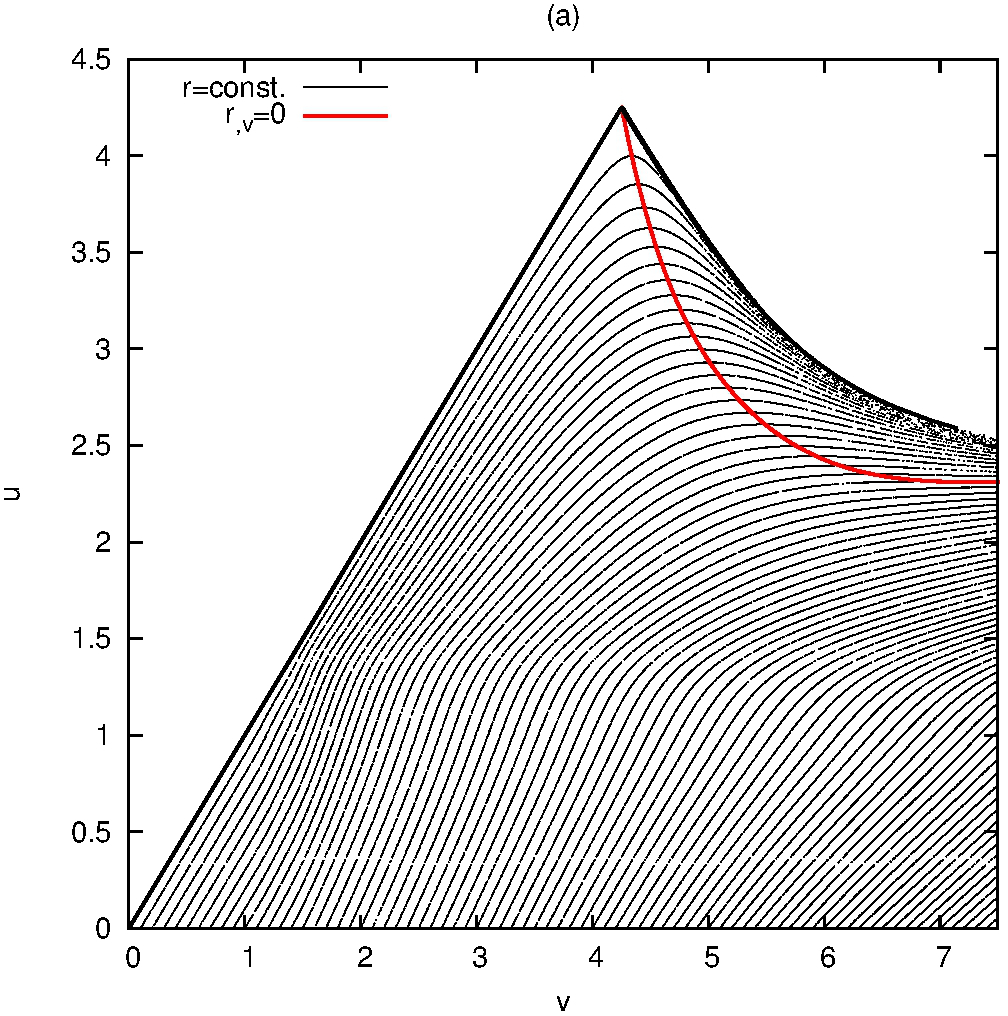}
  \includegraphics[height=.475\columnwidth]{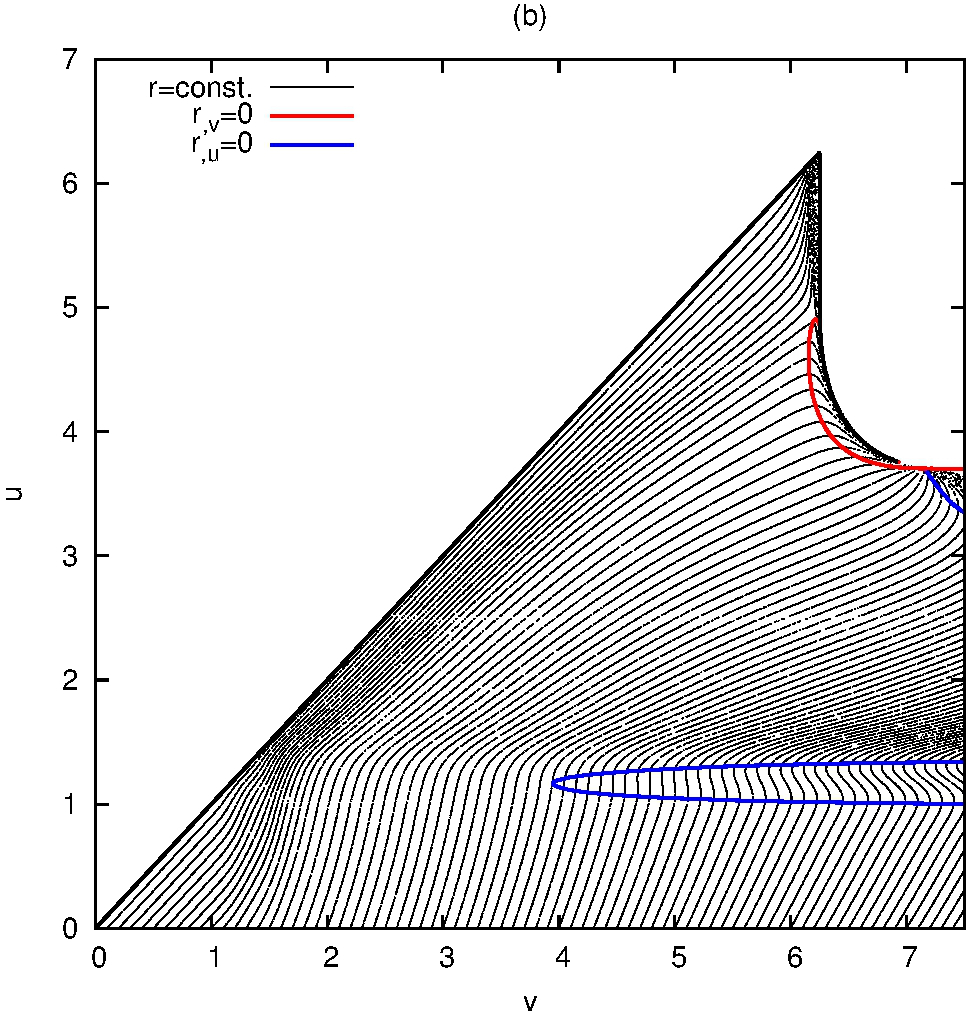}
  \caption{Dynamical singular spacetimes emerging from the $E\overline{MD}$-collapse for (a) $\alpha=-1$ when $\tilde{p}_\phi$ is equal to $0.1$ and~(b) $\alpha=0$ when $\tilde{p}_\phi$ is equal to $0.2$.}
  \label{fig:E-M-D}
\end{figure}

The structures of singular spacetimes, which stem from the $E\overline{MD}$-collapse for $\alpha$ equal to $-1$ and~$0$, are depicted in Fig.\ref{fig:E-M-D}. In both cases the considered process leads to the formation of singular spacetimes for all initial values of dilaton field amplitude. For the former value of dilatonic coupling constant Schwarzschild-type spacetimes emerge, while for the latter naked singularities are observed.

These results confirm the above conclusions concerning the role of particular fields and~couplings in the analysed collapse.



\section{Conclusions}

We examined the results of a~gravitational collapse of electrically charged scalar field in dilaton gravity with a~possibility of phantom coupling of Maxwell and~dilaton fields to gravity. The conclusions regarding the role of particular fields and~couplings in its course are the following:
\begin{itemize}
\item non-phantom dilaton field supports formation of the Schwarzschild-type singular spacetime structures,
\item phantom dilaton field prevents singularity formation due to its strongly repulsive nature,
\item non-phantom Maxwell field 
favours Reissner-\linebreak-Nordstr\"{o}m-type structure formation,
\item phantom Maxwell field supports the creation of the Schwarzschild-type singular spacetime structures,
\item dilatonic coupling constant $\alpha=-1$ diminishes the influence of Maxwell field on the collapse and~thus enhances the role of dilaton field in its course,
\item dilatonic coupling constant $\alpha=0$ reduces the role of dilaton field and~indirectly enlarges the meaning of Maxwell field by acting on electrically charged scalar field.
\end{itemize}





\bibliographystyle{aipproc}   

\bibliography{AIPConfProcMulticosmofun2012}

\IfFileExists{AIPConfProcMulticosmofun2012.bbl}{}
 {\typeout{}
  \typeout{******************************************}
  \typeout{** Please run "bibtex \jobname" to obtain}
  \typeout{** the bibliography and~then re-run LaTeX}
  \typeout{** twice to fix the references!}
  \typeout{******************************************}
  \typeout{}
 }

\end{document}